\documentclass[manuscript, nonacm]{acmart}

\AtBeginDocument{%
  }

\begin{document}

\title{Same Feedback, Different Source: How AI vs. Human Feedback Shapes Learner Engagement}

\author{Caitlin Morris}
\email{camorris@media.mit.edu}
\orcid{0000-0003-3883-2702}
\affiliation{%
  \institution{MIT Media Lab}
  \city{Cambridge}
  \state{Massachusetts}
  \country{USA}
}

\author{Pattie Maes}
\orcid{0000-0002-7722-6038}
\affiliation{%
  \institution{MIT Media Lab}
  \city{Cambridge}
  \state{Massachusetts}
  \country{USA}}
\renewcommand{\shortauthors}{Morris et al.}

\begin{abstract}
When learners receive feedback, what they believe about its source may shape how they engage with it. As AI is used alongside human instructors, understanding these attribution effects is essential for designing effective hybrid AI–human educational systems. We designed a creative coding interface that isolates source attribution while controlling for content: all participants receive identical LLM-generated feedback, but half see it attributed to AI and half to a human teaching assistant (TA). We found two key results. First, perceived feedback source affected engagement: learners in the TA condition spent significantly more time and effort (d = 0.88–1.56) despite receiving identical feedback. Second, perceptions differed: AI-attributed feedback ratings were predicted by prior trust in AI (r = .85), while TA-attributed ratings were predicted by perceived genuineness (r = .65). These findings suggest that feedback source shapes both engagement and evaluation, with implications for hybrid educational system design.
\end{abstract}


\begin{teaserfigure}
  \includegraphics[width=\textwidth]{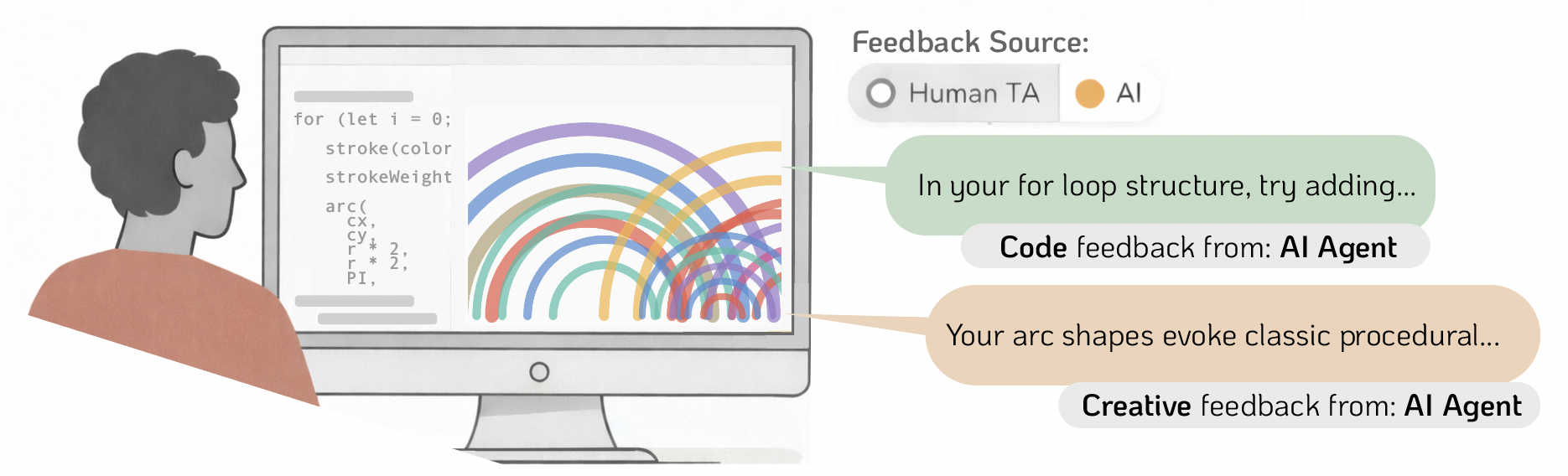}
  \caption{We investigate how feedback source attribution affects learner engagement in creative coding. All participants receive identical LLM-generated feedback on technical and creative dimensions, but half see it attributed to AI and half to a human TA.}
  \Description{Teaser image showing an illustrated student looking at computer code and graphics output on a screen. Text bubbles show snippets of feedback on technical and creative merits, generated by an AI.}
  \label{fig:teaser}
\end{teaserfigure}

\received{20 February 2007}
\received[revised]{12 March 2009}
\received[accepted]{5 June 2009}

\maketitle

\section{Introduction}
Feedback is among the most powerful influences on learning \cite {Hattie2007-zb}. As AI tools increasingly provide feedback in educational settings \cite{Kasneci2023-cs}, understanding how the perceived \emph{source} of feedback shapes learner engagement has implications for how these tools are designed and integrated with human instruction.

Prior research comparing AI and human feedback in education has found no significant overall difference but high unexplained variability \cite{Kaliisa2025-rs}. Two limitations stand out. First, existing studies focus almost exclusively on knowledge acquisition and task performance; these metrics do not reflect effort, engagement, or how learners respond to feedback. Second, no studies control for identical feedback across conditions: all compare AI-generated feedback to human-generated feedback, confounding source with inherent content differences. This makes it impossible to isolate how \emph{believing} feedback comes from AI versus a human shapes learner behavior.

Research that has controlled for content reveals striking attribution effects. Lipnevich and Smith found that college students rated identical essay feedback as more accurate when attributed to an instructor than a computer \cite{Lipnevich2008-eu}. Rubin et al. demonstrated that AI-generated empathic responses were perceived as less empathic, less supportive, and less emotionally satisfying when labeled as AI rather than human, even though the content was identical \cite{Rubin2025-lr}. Social presence theory suggests that learners’ perceptions of an attentive social entity - even through minimal cues such as text-based interaction - can shape effort and engagement \cite{Lowenthal2010-xm,Gunawardena1997-gc,Walther1996-oc, Bandura1977-so}. Perceived human connection in online learning affects motivation, persistence, and depth of processing \cite{Garrison1999-ly}. If learners perceive AI feedback as lacking human presence, they may engage with it differently even when the content is identical to human feedback.

% Vygotsky's Zone of Proximal Development emphasizes that learning is fundamentally social, occurring through interaction with others who provide scaffolding \cite{Vygotsky1979-nz}. Bandura's Social Learning Theory demonstrates that perceived social context shapes not just what people learn but how much effort they invest \cite{Bandura1977-so}. 

\subsection{Experimental Study: Same Content, Different Framing}

 We extend the existing work by examining not just perception but behavioral engagement. Do learners invest more effort when they believe a human is reviewing their work? We situate our study in creative coding, a domain that combines technical and creative assessment on the same artifact. When a learner writes p5.js code to create a visual scene, they exercise technical skill (syntax, function usage, code structure) and creative judgment (color, composition, expression) simultaneously \cite{Processing-Foundation2014-hv}. This lets us additionally ask whether attribution effects differ for technical versus creative feedback.

We designed a creative coding learning interface that isolates attribution effects by holding feedback content constant. All participants receive feedback generated by the same LLM (Claude Sonnet 4) using identical prompting - but half are told it comes from an AI, while half are told it comes from an online human teaching assistant (TA). This design lets us ask clearly: does source attribution affect learner engagement and creative output? And does it affect responses to technical and creative feedback differently? The answers have implications for how hybrid solutions might be designed to maximize learner receptivity and creative growth.

Our research questions are as follows:
\begin{itemize}
\item RQ1: When feedback content is held constant, does source attribution (AI vs. human) affect learner engagement, effort, and creative output?
\item RQ2: Does source attribution interact with feedback type - affecting how learners respond to technical versus creative guidance differently?
\end{itemize}

% \subsection{Contribution and Scope}

% This paper presents early-stage work (current pilot size N=25; target N=60-80) with two contributions:

% \begin{itemize}
% \item Design contribution: A learning interface that isolates source attribution effects by holding feedback content constant—enabling cleaner investigation of how attribution affects engagement and creative output than existing confounded comparisons
% \item Empirical contribution: Evidence that source attribution affects both engagement and what drives perception (prior attitudes for AI, perceived authenticity for TA), though not technical versus creative feedback differentially
% \end{itemize}

\section{Study Design}

\subsection{Learning Interface Overview}

\begin{figure}[h!]
  \centering
  \includegraphics[width=\textwidth]{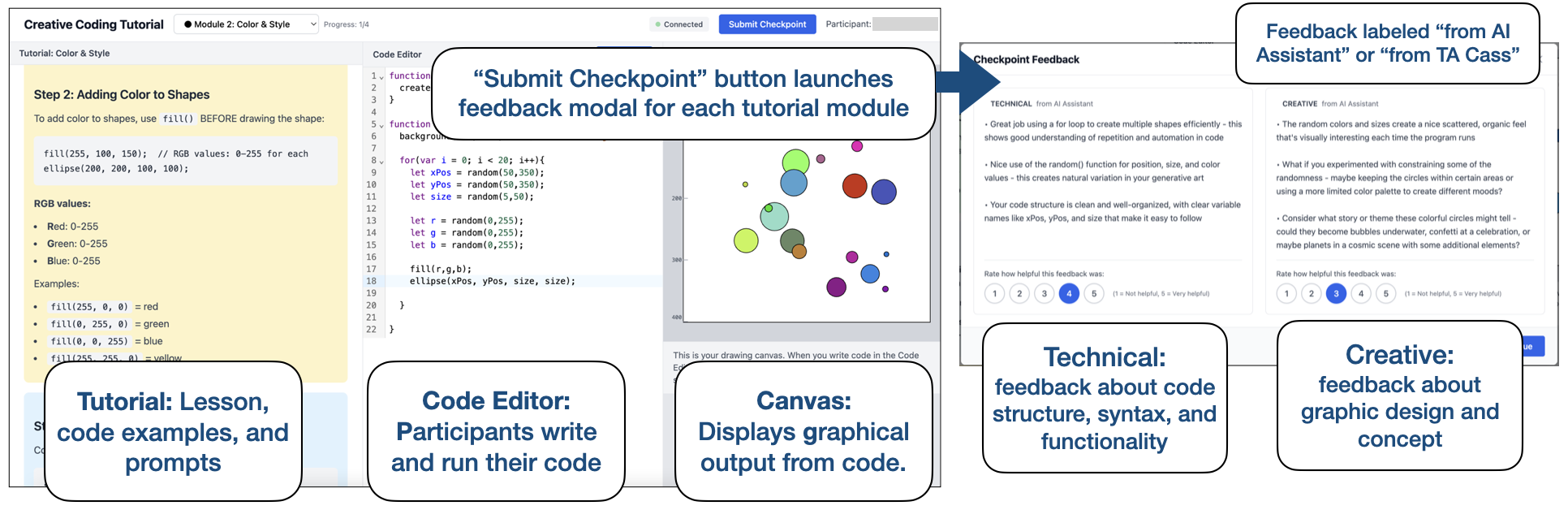}
  \caption{(Left) Web interface for creative coding tutorial. A three-panel design allows simultaneous access to a vertical-scroll tutorial, p5.js code editor, and graphical output canvas. (Right) Checkpoint modal in web interface. Feedback on technical (L) and creative (R) properties are shown simultaneously, attributed to either an AI or TA by research condition. Participants rate each feedback section for usefulness before proceeding to the following module.}
  \Description{A screenshot of a web browser with an interface for reading a code learning tutorial, writing code, and viewing the graphical output.}
    \label{code-interface}
\end{figure}

We developed a web-based platform (Fig. \ref{code-interface}) teaching creative coding through four progressive p5.js modules (shapes, color, interaction, mini-project). Each module takes 10-15 minutes, with the full study lasting approximately 60 minutes.

At each module checkpoint, learners submit their code and receive feedback in two explicitly separated components: \emph{Technical feedback} on code structure, function usage, and programming practices, and \emph{Creative feedback} on visual design and composition. Learners rate each feedback type immediately on 5-point helpfulness scales before proceeding (Fig. \ref{code-interface}).

\subsection{Experimental Design}

This research was approved by institutional IRB and pre-registered on OSF (ID hidden for review). We use a 2×2 mixed design:

\begin{itemize}
\item Between-subjects: Feedback source (AI vs. Human TA)
\item Within-subjects: Feedback type (Technical vs. Creative)
\end{itemize}

\begin{figure}[h]
  \centering
  \includegraphics[width=0.9\linewidth]{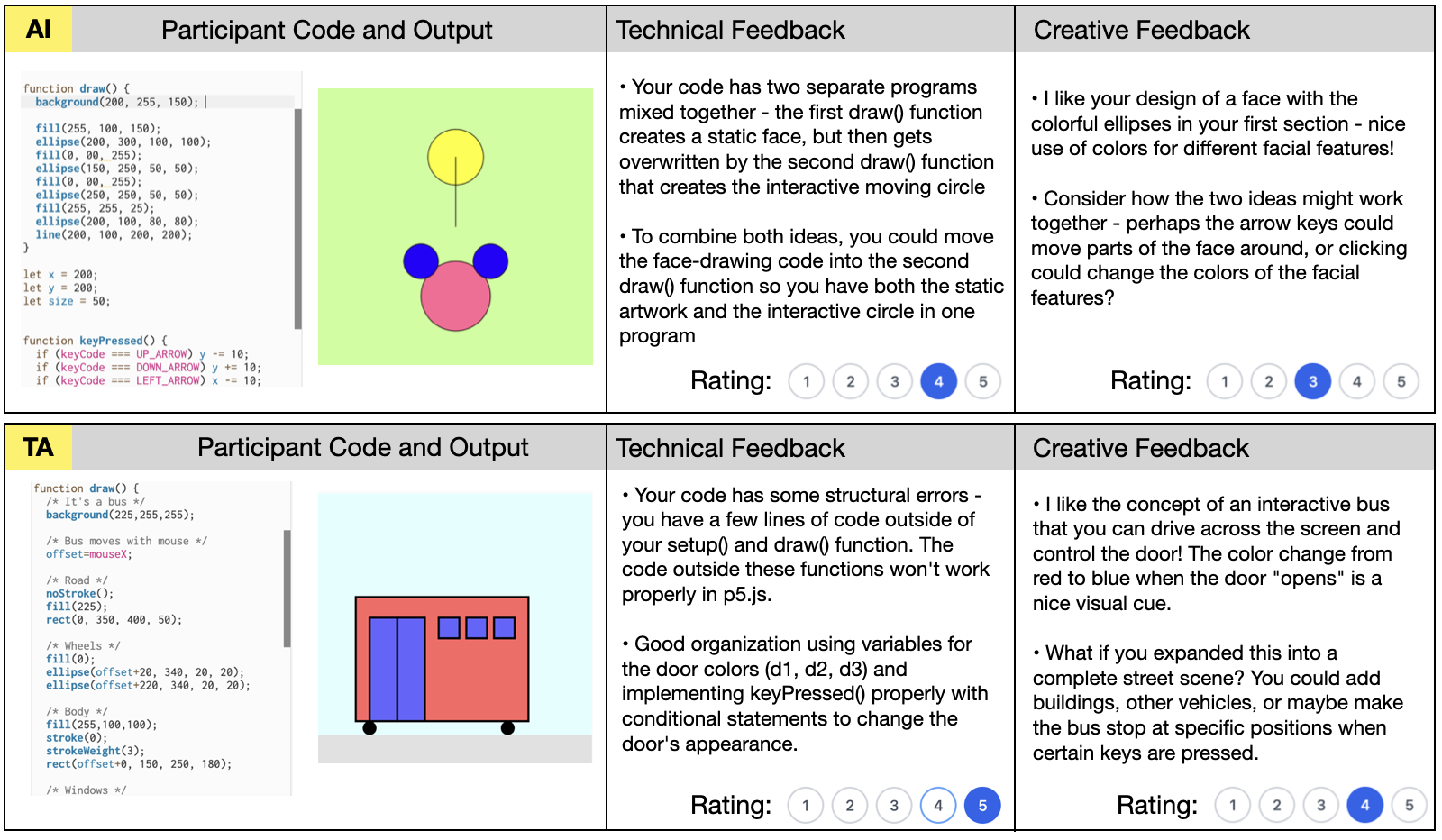}
  \caption{Sample code, graphical output, and feedback from participants in the AI (top) and Human TA (bottom) feedback conditions. From L-R: Code subset, visual output of a single module's exercise, technical feedback and associated rating from the participant, creative feedback and associated rating from the participant.}
  \Description{A composite image showing two rows corresponding to participants in the AI and TA condition.}
  \label{code-examples}
\end{figure}

All participants receive both technical and creative feedback at each checkpoint and rate each type separately, enabling direct comparison of how the same learner perceives these different feedback types from their assigned source.
%%Add note about intro to TA vs AI - No personal information given about "Cass", just a loading screen / pre-scripted intro
Critically, both conditions receive feedback generated by the same LLM (Claude Sonnet 4) using identical prompting structure. The feedback is tailored to each participant's actual code submission in both conditions (Fig. \ref{code-examples}). The only differences are: \emph{Framing:} "AI-Generated Feedback" header vs. "Your TA: Cass" header; and \emph{Timing:} Instant (2-5 seconds) vs. simulated human review delay (~2 minutes with progress indicators e.g. "Your TA is online / Your TA is reviewing your code").

This design isolates the effect of source and timing on feedback perception, while holding feedback content and quality constant. The timing difference reflects the reality that AI feedback is typically instant while human review takes time, supporting perceived authenticity. In both conditions, source attribution was explicit: participants were informed upfront that feedback would come either from an AI agent or from an online teaching assistant, with the TA condition presented via a brief matching screen and introductory message before the first checkpoint.

\subsection{Participants}

For this pilot, we report data from N = 25 participants recruited via Prolific using a two-stage process. A screening survey excluded respondents with no interest in coding or tutorial-based learning (26\% excluded). Participants ranged in age from 20–77 years (M = 41.3, SD = 13.0; median = 39); 18 identified as male and 7 as female. Prior coding experience ranged from none (n = 13), to some experience (n = 7), to substantial experience (n = 5).

Pre-screening also assessed prior AI tool use: 2 participants (8\%) reported never using AI tools, 5 (20\%) rarely, 6 (24\%) sometimes, 8 (32\%) often, and 4 (16\%) very frequently. Participants completed informed consent, worked through the self-paced tutorial with feedback at four checkpoints, and completed a post-study survey. Participants in the TA condition were debriefed regarding the simulated human feedback.

\section{Measures}

We captured both behavioral and perceptual measures. Behavioral engagement and creative output were captured via time spent per module, code execution frequency, lines of code written, reference-checking behavior (click-away events), and deviations from starter code (e.g., additional functions, syntactic variety, and visual complexity). For feedback perception, learners rated technical and creative feedback helpfulness separately at each checkpoint (1-5 Likert). Post-survey items assessed perceived feedback influence and authenticity, AI attitudes (trust for technical vs. creative tasks), and learning outcomes.

\section{Results}

\subsection{Engagement and Time on Task}

\begin{figure}[H]
  \centering
  \includegraphics[width=0.8\linewidth]{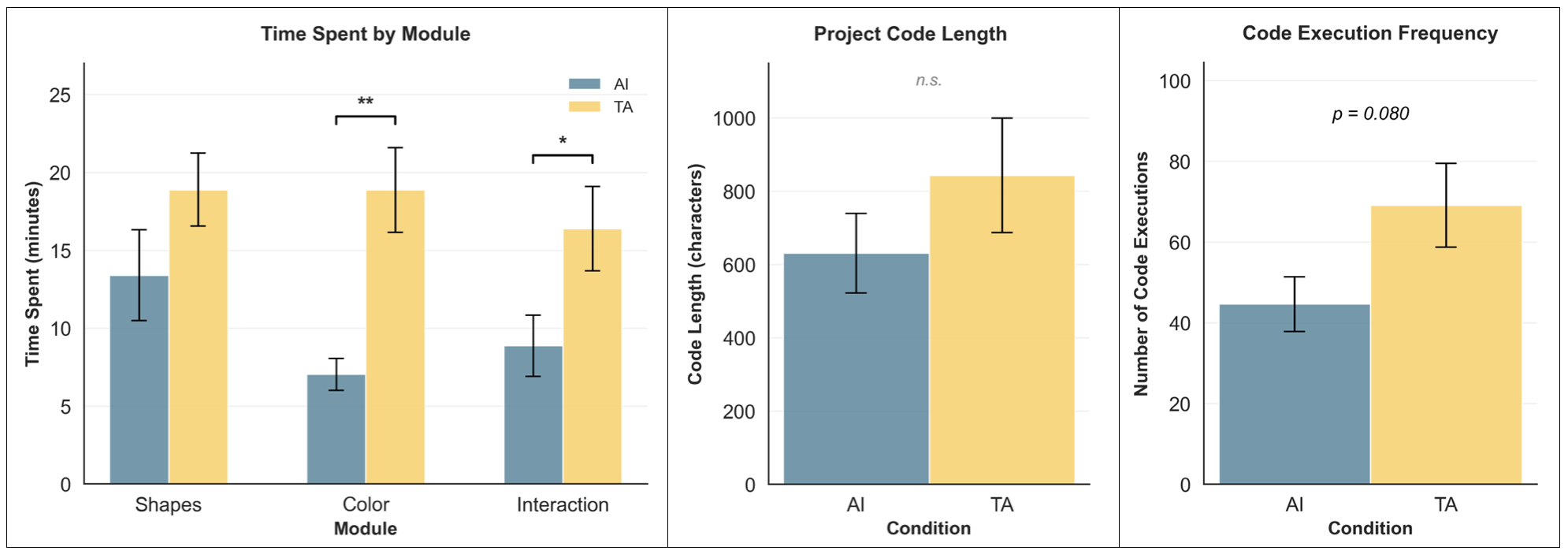}
  \caption{Behavioral engagement by condition. Left: Time spent per module (minutes), with significant differences on modules occurring after initial feedback. Center: Total code length in final project (characters). Right: Code execution frequency across the tutorial.}
  \Description{Statistical figures showing behavioral engagement.}
  \label{behavioral-results-figure}
\end{figure}

TA condition participants showed significantly higher engagement despite receiving identical feedback (Fig. \ref{behavioral-results-figure}). Time on task differences were substantial: on the color module, AI participants averaged 7.0 minutes versus 18.9 minutes for TA participants (t(23)=-3.68, p=.001, d=1.56); on the interaction module, AI averaged 8.9 minutes versus 16.4 minutes for TA (t(23)=-2.14, p=.043, d=0.88). Notably, these effects were larger in later modules (after participants had received at least one round of feedback) suggesting the framing shaped ongoing engagement rather than just initial behavior.

Code execution frequency showed a large effect approaching significance: TA participants ran their code more frequently (M=69.1, SD=37.5) than AI participants (M=44.6, SD=21.4; t(23)=-1.84, p=.080, d=0.80), indicating additional testing and iteration. Additional behavioral indicators showed consistent medium-sized effects: total tutorial time (d=0.47) and click-away events indicating reference-checking (d=0.68). Length of code written showed a similar pattern: TA participants wrote more code (M=843, SD=584) than AI participants (M=631, SD=359; d=0.44).

\subsection{Technical vs. Creative Feedback Ratings}

We found no significant interaction between source attribution and feedback type (F(1,23)=0.95, p=.339). Technical and creative feedback were rated similarly in both conditions (Technical: M=4.23; Creative: M=4.16). There was also no main effect of source (F(1,23)=0.13, p=.723) or feedback type (F(1,23)=0.39, p=.537). Learners did not differentially value technical versus creative feedback based on perceived source. This null finding is notable given prevalent narratives that AI excels at objective/technical assessment while humans are better suited for subjective/creative judgment. Instead, the behavioral differences we observed suggest that social presence effects on engagement may supersede any content-specific beliefs about AI versus human competence.

\subsection{Predictors of Feedback Ratings}
\begin{figure}[h!]
  \centering
  \includegraphics[width=0.7\linewidth]{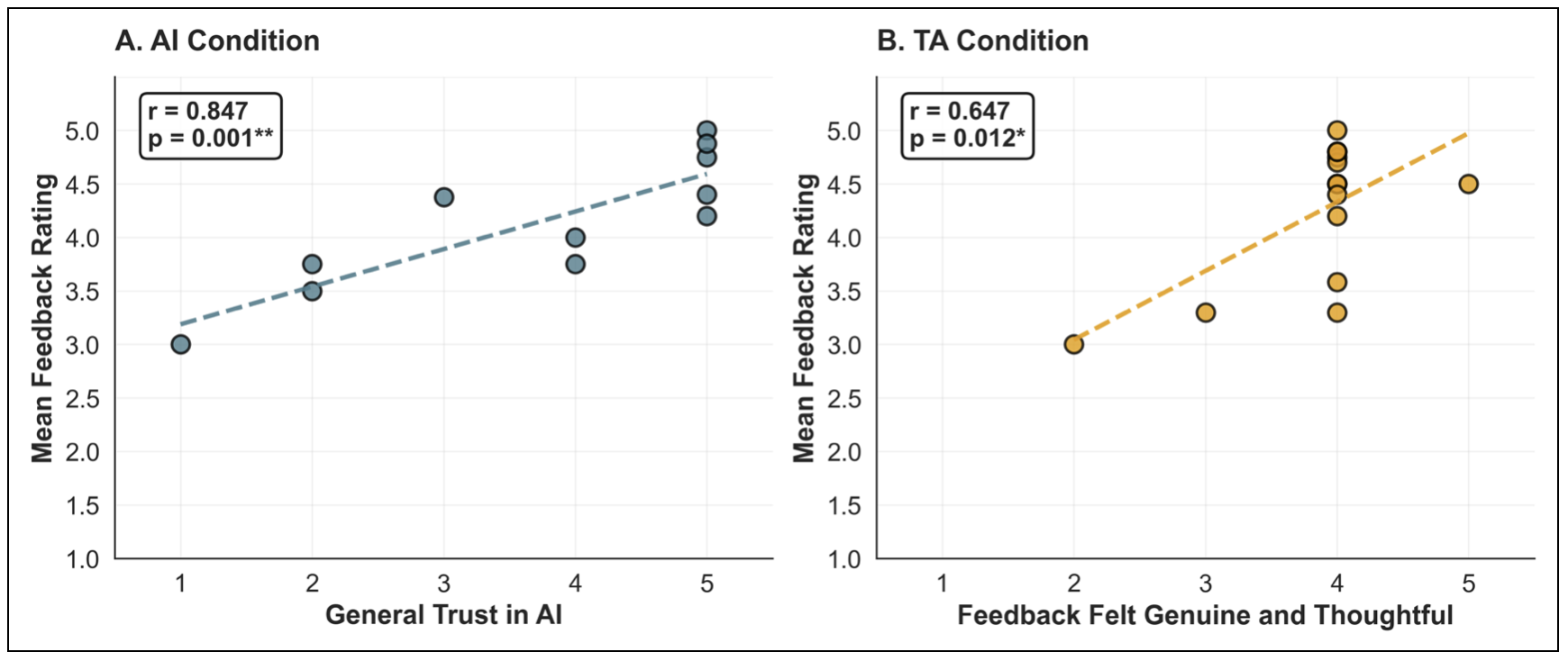}
  \caption{Predictors of feedback ratings by condition. (Left) In the AI condition, prior trust in AI strongly predicted ratings. (Right) In the TA condition, perceived genuineness of feedback predicted ratings. Each point represents one participant's mean feedback rating.}
  \Description{Figure showing predictors of feedback ratings by condition.}
  \label{results-moderators}
\end{figure}

Ratings in each condition were moderated by significantly different factors (Fig. \ref{results-moderators}). In the AI Condition, ratings are primarily driven by prior attitudes toward AI., e.g. trusting AI for creative tasks (r=.847, p=.001) or trusting AI for technical tasks (r=.761, p=.007). Perceived genuineness of the feedback was not a significant predictor (r=.216, p=.523). 
In contrast, in the Human TA condition, the major driver of ratings was perceived authenticity: feeling that the feedback felt genuine (r=.647, p=.012) and that the TA understood their goal with their projects (r=.653, p=.011). Attitudes toward AI were not significant for TA condition participants. 

\section{Discussion}

%%controlled content / constant content?
Our controlled-content design reveals that source attribution shapes both behavior and perception. TA participants invested significantly more time and effort despite receiving identical feedback, an effect that grew stronger across modules after more feedback. Meanwhile, ratings were driven by different factors in each condition: prior AI attitudes predicted AI-attributed ratings, while perceived genuineness predicted TA-attributed ratings. Notably, high ratings in the AI condition did not translate to greater effort: believing feedback is helpful is not the same as being motivated to engage deeply with it. This extends prior work on attribution effects, which focused primarily on perception, by demonstrating that source attribution shapes not just how learners rate feedback but how much they engage with the material. This dissociation aligns with social presence theory, which suggests that even minimal cues of human attention can shape motivation and effort in online environments \cite{Walther1996-oc, Gunawardena1997-gc, Lowenthal2010-xm}. Learners may engage differently when they perceive a human presence attending to their work, even a stranger, due to social motivations that AI does not elicit. These findings suggest that hybrid solutions cannot assume high-quality AI feedback will motivate the same depth of engagement as human feedback, even when learners rate it positively.

\subsection{Limitations}
This pilot (N=25) has limited statistical power, and results should be interpreted cautiously pending the full future sample. The single-session design with four checkpoints cannot capture how attribution effects might shift with longer repeated exposure. The TA condition involves deception, not real human feedback, though the strong genuineness-rating correlation suggests participants believed the manipulation.

\section{Conclusion}
This work contributes a controlled-content methodology for isolating attribution effects in AI-assisted learning, revealing that source framing shapes not just how learners perceive feedback but how much effort they invest. As AI feedback tools become common in education, understanding when and why learners engage differently with identical content based on believed source has implications for how we design, deploy, and frame these systems. 

%%\begin{acks}
%%\end{acks}

\bibliographystyle{ACM-Reference-Format}
\bibliography{paperpile}

\end{document}